\documentclass[english,11pt]{revtex4-2}
\usepackage[latin9]{inputenc}
\usepackage{amsmath}
\usepackage{graphicx}

\begin{document}
\title{Scalable semi-classical implementation of Shor factoring using time-multiplexed
degenerate optical parametric oscillators}
\author{Minghui Li}
\affiliation{Institute of Applied Physics and Materials Engineering, University
of Macau, Macau S.A.R., China}
\author{Wei Wang}
\affiliation{Institute of Applied Physics and Materials Engineering, University
of Macau, Macau S.A.R., China}
\author{Zikang Tang}
\affiliation{Institute of Applied Physics and Materials Engineering, University
of Macau, Macau S.A.R., China}
\author{Hou Ian}
\affiliation{Institute of Applied Physics and Materials Engineering, University
of Macau, Macau S.A.R., China}
\email{houian@um.edu.mo}

\begin{abstract}
A scheme to encode arbitrarily long integer pairs on degenerate optical
parametric oscillations multiplexed in time is proposed. The classical
entanglement between the polarization directions and the phases of
the oscillating pulses, regarded as two computational registers, furnishes
the integer correlations within each pair. We show the major algorithmic
steps, modular exponentiation and discrete Fourier transform, of Shor's
quantum factoring algorithm can be executed in the registers as pulse
interferences under the assistance of external logics. The factoring
algorithm is thus rendered equivalent to a semi-classical optical-path
implementation that is scalable and decoherence-free. The sought-after
multiplicative order, from which the prime factors are deduced, is
identified from a two-dimensional fringe image generated by four-hole
interference measured at the end of the path.
\end{abstract}
\maketitle

\section{Introduction}

Since the now renowned quantum integer factoring algorithm was proposed
by Shor more than two decades ago~\citep{Shor}, experimental implementations
of its compiled versions have been demonstrated on various platforms,
include those on nuclear magnetic resonance (NMR) systems~\citep{Vandersypen},
photonic systems~\citep{Lu,Lanyon}, superconducting circuits~\citep{Lucero},
and ion-trap systems~\citep{Monz}. Despite these demonstrations,
a scalable quantum computer that runs Shor's algorithm still remains
a distant goal that challenges many researchers in the field.

For instance, the scalability challenge for an NMR system is posed
by the fact that the qubits are encoded in the mixed states of magnetic
spins within one molecule, making the system detectable only through
an exponentially scaling number of measurements on the entire ensemble
of the spins~\citep{Schmidt-Kaler}. Using photons as qubits, recycling
techniques has helped to reduce the $\mathit{n}$ qubits needed by
the control register to a single qubit~\citep{Mart=0000EDn-L=0000F3pez},
but the number of qubits in the work register remains the same. Furthermore,
due to low brightness and low collection efficiency, scaling photon
entanglement is still challenging, although great efforts have led
to an ever greater demonstrated number of entangled photons in recent
years~\citep{Wang-1,Wang-2}. In contrast, forming qubits in ion-trap
systems or superconducting circuits is lesser of a problem~\citep{Monz,Barends}.
Nevertheless, other experimental difficulties arise. For ions trapped
inside optical or magnetic wells, qubit control is realized through
focused laser pulses, making the coupling efficiency depend on the
efficacy of fine focusing. The recent invent of on-chip ion traps
has partially remedied this problem~\citep{Romaszko20,Todaro21}.
For superconducting circuits, limited coherence times of the superconducting
qubits, though current technologies has pushed them well beyond the
order of $\mu$s~\citep{Rigetti}, mandates the use of quantum error
correction for scaling~\citep{Koch}. The error-correcting protocols
in turn demand the employment of auxiliary qubits, which spatially
complexify the circuit design~\citep{Kjaergaard,Bharti}.

Moreover, despite the experimental verifications of Shor's algorithm
across diversified physical systems discussed above~\citep{Vandersypen,Lu,Lanyon,Lucero,Monz},
the key algorithmic step of computing modular exponentiation often
still has to rely on external resources when the multiplicative order
is long~\citep{Smolin}. The title of a full-fledged integer-factoring
quantum computer is yet to be claimed. While anticipating the solutions
to the obstacles discussed, we propose here a semi-classical implementation
of Shor's algorithm where the probabilistic nature of full quantum
systems is removed but the characteristics of entangled states retained,
using classical entanglement.

In optics, quantum entangled states are established between a pair
of photons when detections are made by on photodiodes; whereas, classical
entangled states exist among two degrees of freedom within a single
macroscopic light beam~\citep{Spreeuw,Qian}, detectable by interference
fringe patterns~\citep{You}. The inseparable correlations are established
classically between the polarization direction and either the amplitude~\citep{Qian},
or the spatial parity~\citep{Kagalwala}, or the position~\citep{Aiello}
of a light beam. Given the inseparation, violations of Bell-like inequalities
have been demonstrated~\citep{Qian15,Gonzales}. On the theoretical
side, Schmidt analysis intended for quantifying pure-state entanglement~\citep{Eberly}
is generalized to classical fields~\citep{Qian,You}. These studies
suggest that, stripping away the non-locality inferred by the Einstein-Podolsky-Rosen
experiment, classical entanglement plays an substituting role in indicating
local non-classical correlation. Free from decoherence, the long-life
classical analogue is also proved to be detrimental to implementing
the Deutsch algorithm~\citep{Perez-Garcia}, quantum walk~\citep{Goyal},
and the quantum Fourier transform~\citep{Song} without a quantum
system.

Our proposal encodes the classically entangled bits between the phase
states, $\left|0\right\rangle $ and $\left|\pi\right\rangle $, and
the polarization direction states, $\left|H\right\rangle $ (horizontal)
and $\left|V\right\rangle $ (vertical), in degenerate optical parametric
oscillators (DOPOs)~\citep{Loudon}. These oscillators being essentially
phase-coherent single-mode pulses are free from state decoherence
while propagating in fibers, although their intensities need to be
compensated through regular amplification. Their phase coherence and
ability to be time multiplexed~\citep{Wang,Marandi} earn them a
key role in carrying out quantum annealing algorithms in coherent
Ising machine~\citep{McMahon Peter,Inagaki}. Here, we use the time
multiplexity to scale the number of bits necessary to store the computing
data of Shor's algorithm, where the two entangled degrees of freedom
are allocated to, respectively, the control and the work registers.
Though the semi-classical approach does not have concurrence implied
by quantum superposition states, it is scalable and simplifies the
algorithm by eliminating the step of quantum Fourier transform (QFT),
which is implemented by concatenated phase gates. Instead, the QFT
is effectively merged with the last measurement step through interference
pattern detection~\citep{You}, which not only substitutes the projective
measurements for quantum implementation, but also absorbs the transform
process.

In the following, we explain the theoretical model in Sec.~II, demonstrate
the simulation results of the interference patterns that carry computational
significance in Sec.~III. The details of proposed experimental setup
are given in Sec.~III and the conclusions are given in Sec.~IV.

\section{Theoretical model}

The computational bits are encoded in time-multiplexed degenerate
optical parametric oscillators (DOPOs), which are essentially traveling
narrow-width pulses in a free space or fiber cavity. When pumped well
above a power threshold and maintained by a nonlinear crystal such
as a periodically poled lithium niobate (PPLN), each DOPO pulse sustains
a $0$ or $\pi$ phase relative to the pump pulse due to phase sensitive
amplification~\citep{Marandi-1}. These phase states constitute the
first pair of encoding bits $\left|0\right\rangle $ and $\left|1\right\rangle $,
to be used for the control register in the algorithm. Simultaneously,
each DOPO carries a fixed polarization direction, which can be maintained
freely in space or through a polarization-preserving fiber for a fiber
cavity. Thus, the independent horizontal and vertical polarizations
constitute the second pair of encoding bits for the work register.
Throughout their propagation in an optical cavity, particular correlations
among the phase bits and the polarization bits are subsisted without
extra manipulations, allowing entangled computational data to be stored
in the DOPOs.

\begin{figure}
\begin{centering}
\includegraphics[bb=80bp 80bp 680bp 450bp,clip,width=10cm]{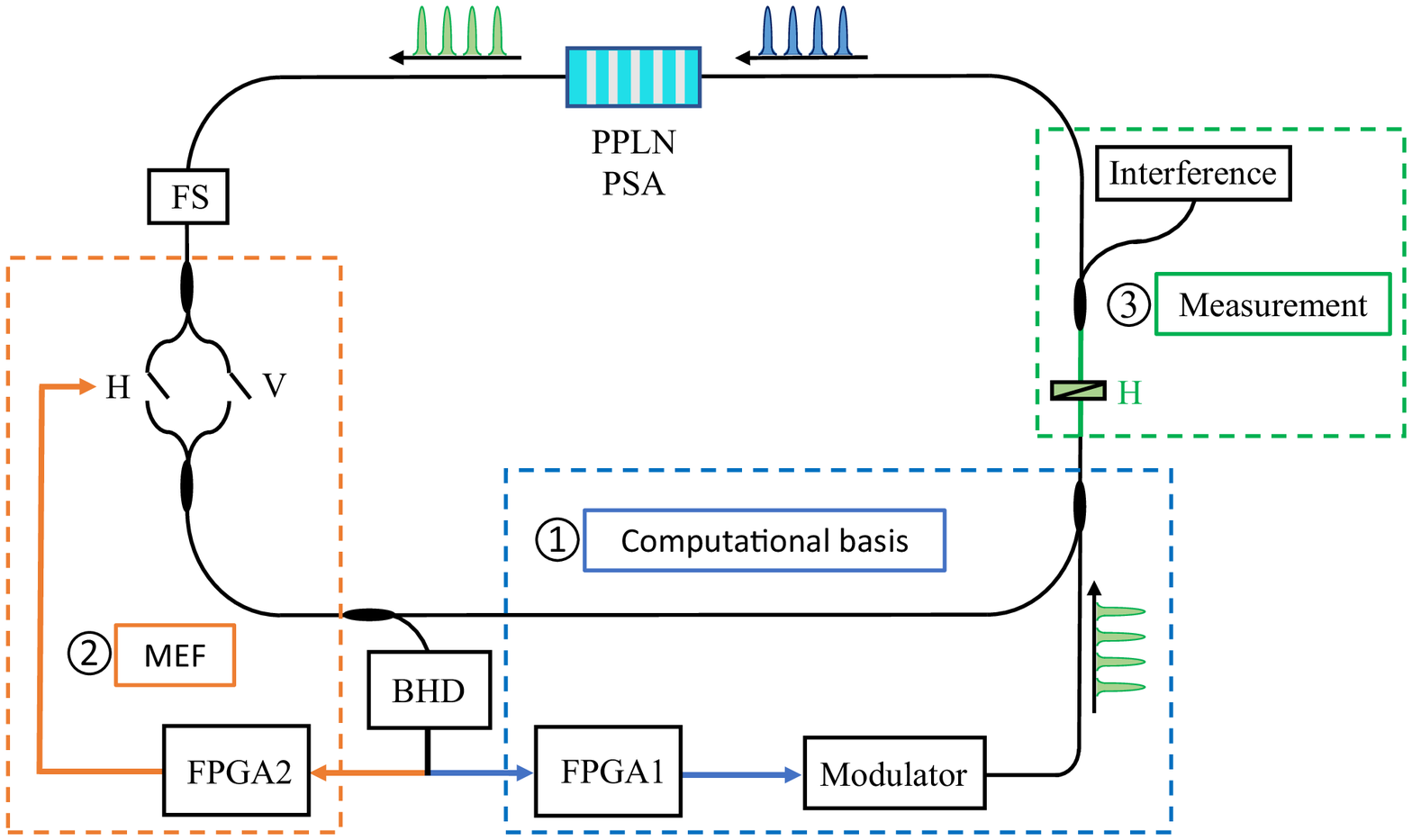}
\par\end{centering}
\caption{Setup design of a classical implementation of Shor's algorithm. PPLN,
periodically poled lithium niobate; FS, fiber stretcher; PSA, phase
sensitive amplification; BHD, balanced homodyne detector; FPGA, field
programmable gate array; Modulator, a module contains the phase modulator
and the intensity modulator. \label{fig:Setup-design1}}
\end{figure}

As shown in Fig.~\ref{fig:Setup-design1}, the computational steps
of Shor's algorithm are accomplished by inserting DOPO manipulating
elements in the optical cavity, with which the information storing
pulses are read out or interfered with correction pulses for data
write-in. Particularly, the modular exponential function (MEF) step
relies on an external field-programmable grid array (the FPGA2 module)
to compute the exponentiation in electrical pulse signals, which are
then fed into the control ports of a pair of optical switches as inverted
signals to produce correlations between phases and polarization directions
of the incident pulses. The QFT has been performed on classical optical
fields using phase modulation of beams (Cf. Ref~\citep{Song}), which
demands costly resources when scaling up. In the proposed setup with
DOPOs here, direct interference measurement on the phases are permitted
without quantum projective collapse, whose fringe patterns implies
the transformation results. Therefore, no particular optical elements
are needed to execute QFT in the optical cavity, as illustrated in
the figure. Before we discuss the details of the experimental design
in Sec. IV, we study how Shor's algorithm can be embedded into the
DOPO-filled cavity below.

To realize the quantum factoring of a composite number $\mathit{N=pq}$,
with $\mathit{p}$ and $\mathit{q}$ are odd primes, one randomly
chooses a base $\mathit{a}$ ($\mathit{\textrm{\ensuremath{0}}<a<N}$)
which is coprime to $\mathit{N}$. Then, one computes the multiplicative
order $\mathit{r}$ of the MEF $f(x)=a^{x}\textrm{mod}N$, which is
the minimum integer that satisfies $a^{r}\textrm{mod}$$N=1.$ For
an even $r$, at least one prime factor is given by the greatest common
divisor (GCD) of $a^{r/2}\pm1$ and $\mathit{N}$. 

The quantum circuits of Shor's algorithm involve at least two registers:
the control register with $n=2\left\lceil \log_{2}N\right\rceil $
qubits to store the numbers $x$ and the work register with $m=$$\left\lceil \log_{2}N\right\rceil $
qubits to store the values of $f(x)$. The control register is initialized
as $|0\rangle$ and the work register is initialized as $|1\rangle$
in computational basis representation. Applying the Hadamard transform
on the control register, it becomes $\mathop{2^{-n/2}\sum_{x=0}^{2^{n}-1}|x\rangle}$,
which is a superposition state of all computational basis states.
Controlled unitary operations for implementing the MEF on the work
register entangle each input value $x$ and the corresponding value
$f(x)$. The entanglement is necessary for speedup in quantum factoring.
Subsequently, the QFT is applied on the control register, which results
in peaks at values of $k2^{n}/r$ (for integer $k$). Finally, the
multiplicative order $r$ can be found by observing the control register.

\begin{figure}
\begin{centering}
\includegraphics[width=10cm]{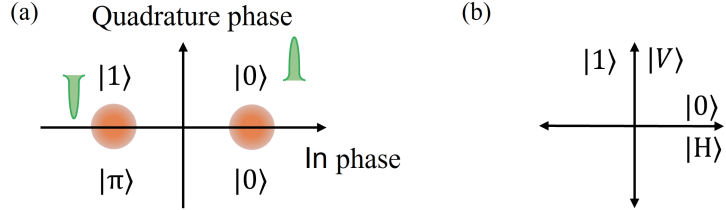}
\par\end{centering}
\caption{Information encoded by the state of a DOPO. (a) Representation of
a classical bit in the control register with the phase state of a
DOPO. (b) Representation of a classical bit in the work register with
the polarization state of a DOPO.~\label{fig:representation}}

\end{figure}

For the general case, we assume that the integer to be factored is
$N$ and the base is $a$. The control register and work register
require $n=2\left\lceil \log_{2}N\right\rceil $ and $m=\left\lceil \log_{2}N\right\rceil $
qubits, respectively. It means we need to prepare computational basis
states from $|0\rangle$ to $|2^{n}-1\rangle$. In order to encode
information in the form of binary numbers, we need physical quantity
with binary values. It has been demonstrated that the state of a DOPO
is a coherent state when pumped above the threshold. The phase of
the coherent state is either $0$ or $\pi$ due to phase sensitive
amplification. For a single DOPO, we use its phase states, $0$ or
$\pi$, to represent the states of a classical bit, $|0\rangle$ or
$|1\rangle$, in the control register (Fig.~\ref{fig:representation}(a)).
Thus, $n$ time-multiplexed DOPOs as a group are required to represent
a $n$-bit number, for example, we can use the phase states of a group
of DOPOs $|0\pi0\pi\cdots0\pi\rangle_{n}$ to represent the binary
number $|0101\cdots01\rangle_{n}$. At the same time, we use the $H$
or $V$ polarization of a DOPO to represent the states of a classical
bit, $|0\rangle$ or $|1\rangle$, in the work register (Fig.~\ref{fig:representation}(b)),
for example, we can use the polarization states of a group of DOPOs
$|HVHV\cdots HV\rangle_{n}$ to represent the binary number $|0101\cdots01\rangle_{n}$.
Since the number of DOPOs is extensible, the experimental scalability
of simulating Shor's algorithm has been realized.

For a DOPO operating above the threshold, its phase state is either
$0$ or $\pi$. We skip the Hadamard transform and prepare an input
initial state straightly. The initial state of the DOPO network is
prepared as:
\begin{center}
\begin{equation}
|\psi_{0}\rangle=\frac{1}{\sqrt{2^{n}}}\sum_{j=0}^{2^{n}-1}|j,0\rangle,\label{eq:initial state}
\end{equation}
\par\end{center}

\noindent which is a product of the superposition of all computational
basis states in the control register and $|0\rangle$ in the work
register.

The implementation of MEF in quantum circuit can be decomposed to
controlled multiplications~\citep{Nielsen}. While feasible in principle,
the number of qubits and quantum gates required is difficult to achieve.
The computation process has been completed in a ``black-box'' fashion
in compiled versions of Shor's algorithm~\citep{Monz}. Here, we
adopt the Montgomery multiplication algorithm~\citep{Montgomery}
on a field-programmable gate array (FPGA) to compute the MEF. The
Montgomery multiplication uses a representation of residue classses
for speedup. For an integer $N$, select an integer $R$ which satisfies
$R>N$ and GCD$\textrm{(}R,N\textrm{)}=1$. Through extended Euclidean
algorithm, we can obtain two integers $R^{-1}$ and $N'$ satisfying
$0<R^{-1}<N$ and $0<N'<R$ and $RR^{-1}-NN'=1$. With these auxiliary
numbers, the Montgomery multiplication computes:
\begin{center}
\begin{equation}
\textrm{Mont}(t_{1},t_{2})=t_{1}\cdot t_{2}\cdot R^{-1}\mod N,\label{eq:MontMulti}
\end{equation}
\par\end{center}

\noindent where $t_{1},t_{2}$ are integers satisfying $t_{1},t_{2}<N$.
When $R$ is taken to be a power of $2$, the Montgomery multiplication
algorithm can avoid trial divisions, which makes it faster than ordinary
modular multiplication~\citep{Koc}. Specifically, the MEF is decomposed
to repeated squaring and modular multiplication by binary method~\citep{Nielsen}:

\begin{equation}
a^{x}\textrm{mod}N=\prod_{j=1}^{n}a^{x_{n-j}2^{n-j}}\mod N.\label{eq:Binary Mod}
\end{equation}
The base $a$ is converted to an $N$-residue for computation, which
is defined as $\bar{a}=aR$ mod $N$. When computing modular multiplication
with Eq.~(\ref{eq:MontMulti}), the modular operation can be replaced
by bit shift and bit-wise AND operations, which are fast in classical
computers. All repeated squaring and modular multiplications are conducted
in the $N$-residue system. By converting the iterative result of
Eq.~(\ref{eq:Binary Mod}) out of the $N$-residue system, the value
of MEF is acquired. The FPGA computes the squaring and multiplication
according to the input value $x$, which is represented by the phase
states of a group of $n$ sequential DOPOs. The values of MEF are
stored in the polarization states of the same group of DOPOs. For
the convenience of the following measurement experiment, we change
the MEF value to $a^{x}(\textrm{mod}N)-1$. The state of the DOPO
network becomes a classical entangled state:
\begin{center}
\begin{equation}
|\psi_{1}\rangle=\frac{1}{\sqrt{2^{n}}}\sum_{j=0}^{2^{n}-1}\left|j,a^{j}\mod N-1\right\rangle .\label{eq:MEF state}
\end{equation}
\par\end{center}

\noindent The degree of entanglement of $|\psi_{1}\rangle$ can be
quantified by the Schmidt number $K$.

In order to find the multiplicative order contained in the control
register, we need to measure the work register first. In general,
$n$ polarizers are required to select a random MEF value represented
by the polarization states of a DOPO group. But for simple demonstration,
we use only one horizontal polarizer to measure the work register
first. As a result, all DOPOs with vertical polarization are eliminated.
Since we treat $n$ sequential DOPOs as a group, only those values
of MEF represented by $|HH\cdots H\rangle_{n}$ polarization state
are kept unchanged. So the state of the DOPO network becomes:

\begin{equation}
|\psi_{2}\rangle=\frac{1}{\sqrt{2^{M}}}\sum_{j=1}^{M}\left|x_{j},0\right\rangle ,\label{eq:Final state-1}
\end{equation}

\noindent where $M$ is the number of input values whose MEF is equal
to 1. Apparently, the state $|\psi_{2}\rangle$ is a product state.
If we can read all the values in the control register, the multiplicative
order can be found. However, for quantum computer, measurement will
lead to collapse of the superposition state in the control register,
which hinders further measurement. Hence, the QFT is applied on the
control register to read out the multiplicative order~\citep{Nielsen}.

Consider for a specific case with $N=15$ and $a=7$ for executing
the algorithm. We have $n=8$ and $m=4$. However, $n=4$ qubits are
enough for the control register due to the multiplicative order $r$
in this situation is $4$~\citep{Lu}. Consequently, a network of
64 DOPOs is required to represent all the computational basis. We
can rewrite $|\psi_{1}\rangle$, with orthonormal basis in phase and
polarization degrees of freedom, in a classical perfectly organized
form~\citep{Qian}:
\begin{align}
|\psi_{1}\rangle= & \frac{1}{2}\left[\frac{1}{2}\left(|00\rangle+|0\pi\rangle+|\pi0\rangle+|\pi\pi\rangle\right)|00\rangle\right]\otimes|HHHH\rangle\nonumber \\
 & +\frac{1}{2}\left[\frac{1}{2}\left(|00\rangle+|0\pi\rangle+|\pi0\rangle+|\pi\pi\rangle\right)|0\pi\rangle\right]\otimes|HVVH\rangle\nonumber \\
 & +\frac{1}{2}\left[\frac{1}{2}\left(|00\rangle+|0\pi\rangle+|\pi0\rangle+|\pi\pi\rangle\right)|\pi0\rangle\right]\otimes|HHVV\rangle\nonumber \\
 & +\frac{1}{2}\left[\frac{1}{2}\left(|00\rangle+|0\pi\rangle+|\pi0\rangle+|\pi\pi\rangle\right)|\pi\pi\rangle\right]\otimes|VVHH\rangle.\label{eq:Classical form}
\end{align}

\begin{figure}
\begin{centering}
\includegraphics[width=10cm]{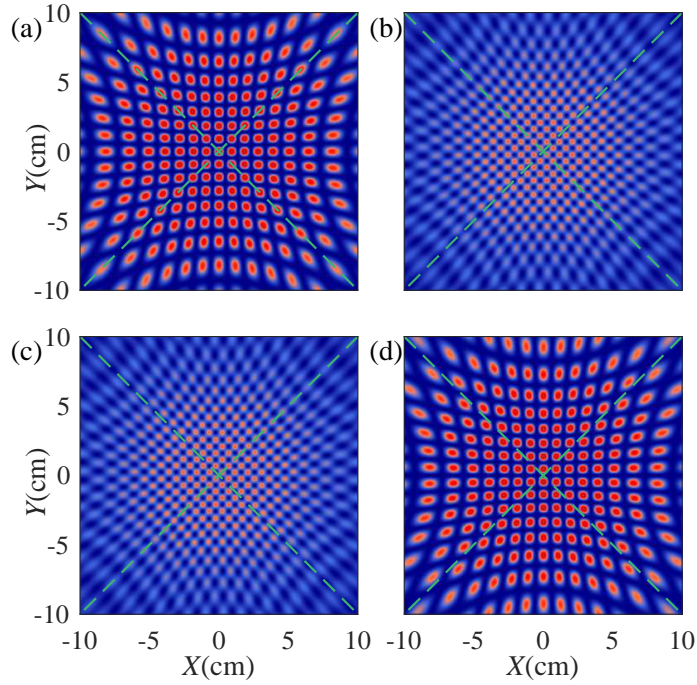}
\par\end{centering}
\caption{Simulation results of measurement through four-hole interference patterns.
The phase states of a group of DOPOs are: (a) $|0000\rangle$, (b)
$|0\pi00\rangle$, (c) $|\pi000\rangle$, (d) $|\pi\pi00\rangle$.~\label{fig:Simulation-results}}

\end{figure}

\noindent So the Schmidt number is $K=1/(4\times2^{-4})=4$ , which
is equal to the multiplicative order of MEF. In particular, the degree
of entanglement will increase with the multiplicative order. It shows
that an extensible degree of entanglement is also necessary for scalable
Shor's algorithm. After measurement of the work register, the state
of the DOPO network becomes:

\begin{equation}
|\psi_{2}\rangle=\frac{1}{2}\left(|0\rangle+|4\rangle+|8\rangle+|12\rangle\right)|0\rangle.\label{eq:Final state}
\end{equation}
Since the measurement will not result in a collapse of the register
state for classical implementation, we can directly measure the control
register. We separate a group of DOPOs into four holes for interference~\citep{You}
through delay lines, so the phase states of $4$ DOPOs can be read
at the same time. Even though some groups only contain two DOPOs,
they can also form interference patterns with stripes. Nevertheless,
those intact groups can form interference patterns with bright or
dark spots. All the interference patterns of these temporally separated
DOPO groups form a video that contains $16$ frames, as shown in the
Supplementary Video. The interference patterns are distinguished through
their symmetry character(Fig.~\ref{fig:Simulation-results}). From
the phase states of a group of DOPOs, we derive four values: $0\textrm{\textrm{,} }4\textrm{, }8\textrm{, and }12\textrm{.}$
The multiplicative order is deduced to be $r=4$. Therefore, the prime
factors of $N$ could be obtained.

\section{Experimental realization}

The experiment setup to simulate a scalable version of Shor's algorithm
is based on a network of time multiplexed DOPOs~\citep{Inagaki}
combined with computation and measurement parts. Fig.~\ref{fig:Setup-design1}
shows the schematic, which mainly contains three parts. The laser
is a continuous wave laser with a wavelength of 1536~nm, which is
suitable for dispersion-shifted fiber. The laser output is split into
two paths, one of which is phase-modulated and used as a local oscillator
for the BHD, and another one is modulated by an intensity modulator
for pulse generation. The generated pulse train with a repetition
multiplicative order of $T_{\mathrm{rep}}$ is further split into
two paths. One path is launched to the modulator module to generate
the feedback pulses according to the feedback signal from FPGA1. The
other is used to synchronously pump the PPLN for phase sensitive amplification
after the second harmonic generation process is completed by another
PPLN.

One DOPO is formed with the PPLN and a fiber-ring cavity. For realizing
a network of time multiplexed DOPOs, we set the round trip time as
$T_{\mathrm{rt}}=60T_{\mathrm{rep}}$. The PSA process is designed
with type 0 phase matching ($e$$\rightarrow\textrm{\ensuremath{e}}+e$),
so that the DOPOs are generated with horizontal polarization. In the
beginning, the optical switch allowing for horizontal polarization
is connected, so that the DOPOs can operate continuously. To maintain
the phase state of each DOPO, the FS is used to stabilize the length
of the cavity. As a result, the phase difference between two pump
pulses of the same signal pulse is compensated by the cavity length,
so that the signal pulse and the pump pulse are synchronous and the
relative phase difference between them is fixed at $0$ or $\pi.$

\begin{figure}
\begin{centering}
\includegraphics[width=14cm]{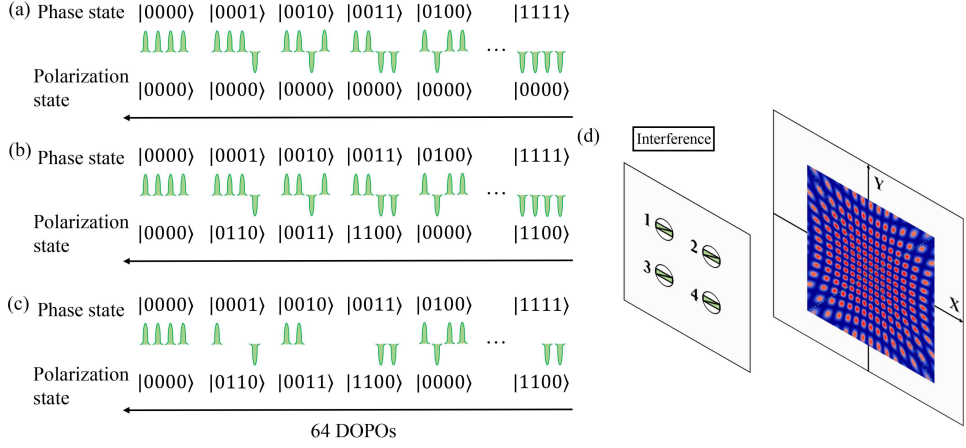}
\par\end{centering}
\caption{The states of DOPOs. (a) Initial state of the DOPO network. (b) State
of the DOPO network after MEF. (c) State of the DOPO network after
measurement of a horizontal polarizer. (d) A general design for measurement.
\label{fig:pulses states}}
\end{figure}

To prepare the computational basis, we use the BHD to measure the
in-phase component $c_{j}$ of DOPOs and the measurement results are
sent to FPGA$1$. The phase state of a DOPO is treated as $0$ or
$\pi$ when $c_{j}$ is positive or negative, respectively. The phase
state of a DOPO is randomly $0$ or $\pi$, so we need to change it
to the set value required for preparing the computational basis. FPGA1
computes the feedback signal $f_{j}=-2(1-\delta_{ij})c_{j}$, where
$\delta_{ij}$ means comparing the phase state of $j\textrm{th}$
DOPO and $\mathit{i}$th set value shown in Fig.~\ref{fig:pulses states}(a).
The $0$ or $\pi$ phase state means the quadrature component of a
DOPO is zero and therefore the coupling pulse only controls the in-phase
component of it. When $f_{i}=-2c_{j}$, the phase of the coupling
pulse is $\pi$ relative to the DOPO, which is realized by phase modulation.
Then, the feedback signal synchronously controls the coupling pulses
through an intensity modulator, so that all the phase states of DOPOs
become the same as the set values. 

The implementation of MEF relies on FPGA2 and two optical switches.
After the superposition state of all computational basis is prepared,
FPGA2 receives the phase states of a group of DOPOs and computes the
MEF through Montgomery multiplication in one round. The values of
MEF are stored in the polarization states of the same group of DOPOs
in the next round. The stored procedure is completed by two optical
switches which allow for horizontal and vertical polarization, respectively.
The state of the DOPO network after MEF is shown in Fig.~\ref{fig:pulses states}(b).

The measurement part contains two steps: first, a polarizer is used
to select all horizontal polarized DOPOs (Fig.~\ref{fig:pulses states}(c));
second, a four-hole interference is exploited to measure the values
represented by a group of DOPOs. The DOPOs are split, through a coupler,
into four paths, which have different delay times so that a group
of DOPOs can arrive at the screen simultaneously. Finally, we can
read the results through interference patterns. A general four-hole
measurement design is shown in Fig.~\ref{fig:pulses states}(d).
Four adjustable polarizers are fixed at four holes, respectively.
The selection of different MEF values could be completed through change
of the polarizers to be horizontal or vertical. Subsequently, the
phase states of a DOPO group are derived through the interference
patterns.

\section{Conclusions}

We have proposed a scalable implementation of Shor's algorithm using
an entirely classical optical system. Specifically, the degrees of
freedoms from the phase and the polarization direction of DOPOs were
used to represent the entangled bits of the algorithmic control and
work registers, making the design free from decoherence innate to
qubit systems. Stored as an order pulse train in a ring cavity, e.g.
an optical fiber, the DOPOs are interfered with oscillating pulses
produced by external optical switches and FPGA logics to execute the
key steps of modular exponentiation and discrete Fourier transform.
The oscillating pulses of fixed width are multiplexed in time while
running along the cavity length. Any bit length needed by the input
integer can therefore be accommodated by sufficient cavity length
and pulse amplification, making the implementation scalable. 

This implementation is semi-classical because the entangled degrees
of freedom permit half-parallel operations on the integers stored.
Manipulations on the phases are immediately reflected in the accompanying
polarization directions although the bits represented by the full
pulse train are operated on one by one in time. We note that the time
multiplexing that scales linearly with the bit length are only hardware
bounded, i.e. the retention conditions of a fiber as the ring cavity.
From the perspective of computation, its scaling is accounted by space
complexity only and not related to query or time complexity. Although
the classical parametric oscillators are unable to form superposition
states and perform projective measurements for full parallelism characteristic
of quantum systems, the major concern over time complexity by Shor
and other factoring algorithms is indeed addressed. Finally, the computed
registers are read by sending the polarized beams of the DOPOs through
a four-hole interference. The fringe patterns are uniquely mapped
to the multiplicative order sought by the algorithm, where the generation
and mapping does not contribute to the algorithmic complexity.

\end{document}